\documentclass[nobibnotes, longbibliography, reprint,aip,rsi]{revtex4-1}
\usepackage{amsmath,amssymb}
\usepackage{graphicx}
\usepackage{color}
\usepackage{url}
\usepackage{hyperref}

\begin{document}

\title{Focus on Imaging Methods in Granular Physics}

\author{Axelle Amon}
\affiliation{Institut de Physique de Rennes, UMR UR1-CNRS 6251, Universit\'e de Rennes 1, 35042 Rennes, France}

\author{Philip Born}
\affiliation{Institut f\"ur Materialphysik im Weltraum, Deutsches Zentrum f\"ur Luft- und Raumfahrt, 51170 Cologne, Germany}

\author{Karen E. Daniels}
\affiliation{Department of Physics, North Carolina State University, Raleigh, NC, 27695 USA}

\author{Joshua A. Dijksman} 
\affiliation{Physical Chemistry and Soft Matter, Wageningen University \& Research, Wageningen, The Netherlands}

\author{Kai Huang}
\affiliation{Experimentalphysik V, Universit\"at Bayreuth, 95440 Bayreuth, Germany}

\author{David Parker}
\affiliation{School of Physics and Astronomy, University of Birmingham, Birmingham B15 2TT, UK}

\author{Matthias Schr\"oter}
\affiliation{Institute for Multiscale Simulation, Friedrich-Alexander-Universit\"at Erlangen-N\"urnberg (FAU), 91052 Erlangen, Germany.}

\author{Ralf Stannarius}
\affiliation{Institut f\"ur Experimentelle Physik, Otto-von-Guericke-Universit\"at, 39106 Magdeburg, Germany}

\author{Andreas Wierschem}
\affiliation{Institute of Fluid Mechanics, Friedrich-Alexander-Universit\"at Erlangen-N\"urnberg (FAU), 91058 Erlangen, Germany}

\date{\today}

\begin{abstract}
 
Granular materials are complex multi-particle ensembles in which macroscopic properties are
largely determined by inter-particle interactions between their numerous constituents. In order to understand
and to predict their macroscopic physical behavior, it is necessary to analyze the composition and interactions at the level of individual contacts and grains. To do so requires the ability to image individual particles and their local configurations to high precision. A variety of complementary imaging techniques have been developed for that task. In this introductory paper accompanying the Focus Issue, we provide an overview of these 
imaging methods and discuss their advantages and
drawbacks, as well as their limits of application.
\end{abstract}

\maketitle
\section{Current Challenges \label{sec:challenges} }

\begin{center}
{\it To see a world in a grain of sand,\\ and a heaven in a wild flower.} 
\end{center}

The poem \textit{Auguries of Innocence} by William Blake illustrates one of the complexities of granular physics: Each grain of sand is unique~\cite{Greenberg2008} and the entirety of particle-particle interactions in a sand pile is unpredictable~\cite{Duran2000}. While walking on a beach, one intermittently experiences the transition between a rigid, solid-like state and a fluid-like state. One leaves behind the stress loading history in the form of footprints~\cite{Geng2001,Herminghaus2013}. Stepping into the water, one recognizes the sediments are looser in comparison to the partially wet
sand on the beach and susceptible to the surrounding fluid flow, leading to sand ripples~\cite{Charru2013}. 

Continuum descriptions based on empirical assumptions can successfully describe rapid flows and sufficiently-dilute granular gases~\cite{Jenkins1983,Goldhirsch2003,MiDi2004}. However, continuum approaches fail to describe slow dense flows or critical behavior such as intermittent flows, jamming and pattern formation~\cite{Liu2010,Aranson2009}. These systems are governed by phenomena which are hard to model in continuum descriptions: strong dissipation at the contacts between the grains due to  friction~\cite{Luding1995}, inelastic deformation~\cite{Brilliantov2004}, and cohesion~\cite{Mueller2016a}, etc. Moreover, the athermal nature of the system does not allow for the use of statistical physics to connect the micro-scale to the macro-scale. 

One of the major issues in modeling granular materials arises from the fact that what happens at the scale of a single grain can impact the response of the whole material. In static piles and dense flows, the distribution of stress is governed by force chains~\cite{Dantu1957}. Those force chains sensitively depend on the individual contacts between the grains, and the history of loading those contacts \cite{Geng2001b,Majmudar2005}. Consequently, local properties of the contacts, such as their typical orientation or the nature of the frictional contacts, can modify the mechanical behavior of the system. For example, anisotropy in the orientation of those contacts~\cite{Radjai1998}, arising due to shear, can have a major impact on the macroscopic response.

On an intermediate scale between the the size of the grains and the sample as a whole, it has been shown that the non-affine motion of the system, at the scale of clusters of typically ten grains, plays a non-negligible role in the mechanical response of the system~\cite{Radjai2002}. This non-affine motion seems to control numerous features of amorphous materials such as the thickness of shear bands~\cite{Schall2010} or the eddy-like structures in dense flows~\cite{Radjai2002}. Several nonlocal effects have been observed, particularly in confined flows. For example, it has been demonstrated that shear bands generate mechanical noise even deep into the seemingly-static phase~\cite{Nichol2010,Reddy2011}.

Therefore, it is of paramount importance for a description of granular matter to be able to make observations on many length scales: from the grain or contact scale, through the mesoscopic effects such as nonlocality and shear banding which appear on the $\sim 10$ grain scale, even up to the scale of a full sample that may contain billions of grains. 

At the grain scale, there are two generic issues making it difficult to extract data with any imaging technique.  First, our most advanced imaging technologies have been developed to act as extensions for our eyes, which operate in the visible spectrum.  However, most granular materials are opaque in this range of wavelengths. Even if the particles were transparent, their refractive index would not generically match that of the most common interstitial fluids (air or water), leading to multiple scattering.  Second, a large volume of raw data is required to analyze a complete granular system: even just a simple sugar cube contains on the order of $10^5$ individual grains. To identify the center of mass and orientation of each of these grains, it is necessary to identify its spatial extent using several thousand voxels (3D pixels). As a consequence,  several gigabytes of data need to be collected to analyze a single static packing of grains. Moreover, trying to describe any kind of dynamics will compound the problem by requiring a sufficiently high frame rate to collect that data.  

There are a number of ways to address opacity issue, even with visible light. A common technique is to perform quasi-two-dimensional (Q2D) experiments, allowing for the complete tracking of particle positions \cite{Reis2006,Huang2015} as well as the particle-particle interactions \cite{daniels:17}. Another option is to restrict the data collection to the surface of the granular system, with the disadvantage that bulk properties can differ significantly from the behavior at the surface 
\cite{zhang:06,orpe:07,jerkins:08,desmond:09,Tennakoon1999,Raihane2011}.
If the volume fraction of the particles is sufficiently low, stereo-camera or volumetric methods can be used to capture three-dimensional (3D) dynamics~\cite{adrian2011particle, raffel2007particle, cierpka2012particle}. If the particles are optically-transparent, there exist two additional options. Where the interstitial liquid can be chosen freely, index matching provides a means to acquire 3D data~\cite{dijksman:17}.  Alternatively, it is possible to embrace the multiple scattering effects, and use coherent light to gather information from the resulting speckle pattern~\cite{amon:17}. 
Finally,  we can abandon the range of visible light entirely, and instead use penetrating radiation to obtain data from the bulk of the sample. A variety of such methods are extensively covered in this focus edition, covering terahertz electromagnetic radiation~\cite{born:17}, radar~\cite{huang:17}, positron emission~\cite{parker:17}, nuclear magnetic resonance~\cite{stannarius:17}, and  X-ray tomography~\cite{weis:17}.

Several solutions also exist for the data-bandwidth issue, which are common to the various methods. Limiting the analysis to 2D or the surface of a 3D system is an effective way to decrease the number of particles under analysis, at the expense of ignoring the bulk behavior. To access the bulk, it is possible to restrict the analysis to quasi-static systems in which the driving is stopped during each cycle to take a snapshot. This will reduce the necessary frame rate to zero, eliminating problems with bandwidth.  Alternatively, the experiment can be constructed so that only a subset of tracer particles are visible to the image acquisition system, or take depth-averaged signals, in order to retain the dynamics.

The present focus issue mainly covers imaging with electromagnetic waves. Nevertheless, acoustic waves can also be used to measure local velocities of particles in dilute flows~\cite{Takeda1986,Manneville2004} or elastic properties of the effective medium for dense systems~\cite{Jia2004}. A problem is that multiple scattering prevents spatial resolution with acoustic waves in dense granular piles. Another difficulty when using acoustic waves is the intrusiveness of the probe that acts at the level of the contacts between the grains.

This overview article, as well as the following focus issue on imaging granular particles, aims to provide guidance and orientation concerning the experimental techniques which help to face all these challenges.

\section{Acquiring particle positions, orientations, \& shapes \label{sec:positions}}

\begin{figure*}[htbp]
\centering
\includegraphics[width=\textwidth]{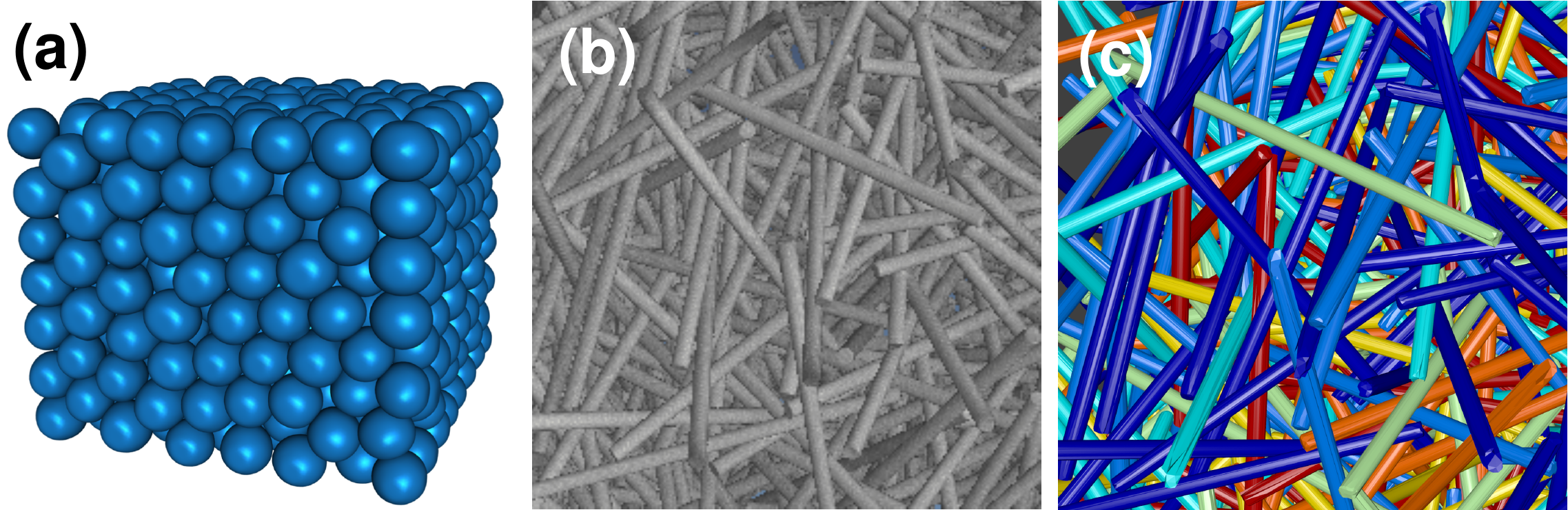}
\caption{3D imaging of granular particles. 
(a) Compressed hydrogel spheres with 2 cm diameter(Educational Innovations), imaged by laser sheet scanning in an index-matched liquid. Particle deformations in a compressed packing of soft spheres are clearly visible, for example in the flattening of spheres at the top. Such deformations are measurable at each particle-particle contact and yield information about contact forces. 
(b) Rendering of the volume data obtained by X-ray tomography, the particles are cylinders with 40 mm length and 1.4 mm diameter (Spaghettini Barilla N.~3). An animated version of similar raw data can be found at \cite{youtube}. (c) Rendering of the same volume as in the last panel, but after the detection of all cylinder positions and orientations. The particles are color coded with their contact number. Images b and c by courtesy of Cyprian Lewandowski.
\label{fig:sec_2}
}
\end{figure*} 

To acquire particle positions, orientations and shapes, a two-step process is necessary.  First, an image is made of a particle and its immediate surroundings. Second, the required particle information is extracted through data analysis of those images. For both steps, there are multiple methods to chose from; in this Focus Issue, we discuss both steps in some detail. 

To image a particle, some contrast between the particle and their surrounding medium is required. The most obvious method to detect particles is to use the visible part of the electromagnetic spectrum. In this range, it suffices to use standard cameras to obtain digital images of a collection of particles. The only difference between 2D and 3D imaging is whether each configuration is represented as a single 2D image or as a stack 2D slices. Due to both absorption and scatter, however, the visible spectrum has limited penetration into most materials. Since the time of the first packing experiments by Bernal~\cite{bernal1960}, one solution has been to create 3D images by physically disassembling the packing and taking an image at each step. This method is still used in the present day for for sufficiently slow flows~\cite{2004_prl_fenistein}.

To avoid such destructive methods, and considerably improve the data collection rate, modern experiments commonly use transparent granular media. For optical transmission through a pile of transparent glass marbles, however, scatter remains a significant limitation in going deeper than a few particles inward from the boundary. The solution is to reduce the scatter by  immersing the solid particles in an index-matched medium.  To obtain the necessary image contrast, at least one of the two phases must be stained with a fluorescent dye. Thus, by illuminating a cross section of the medium with a sheet of light (usually via a laser), the fluorescent response of the dyed material within the sheet is captured by the camera.  By moving the light sheet with respect to the sample and recording a series of slices, a 3D image of the medium can be created. 

In the resulting  3D image, computerized-post processing algorithms can then be used to track particles, measure flow velocities, or identify shapes. This technique of Refractive Index Matched Scanning (RIMS) has been covered by several review articles~\cite{wiederseiner, 2011_revsciinstr_dijksman}. The RIMS article in this Focus Issue will explore the application of RIMS~\cite{dijksman:17} to hydrogel particles in particular. These are soft elastic solids with a refractive index close to that of water. The softness of hydrogels provides a key feature: they deform significantly under modest loads. The presence of deformations at each contact  makes it possible to quantify individual contact forces on hydrogels, as is discussed in Sec.~\ref{sec:forces}. The central challenge is to find the nonspherical contour of the particle. An example can be seen in Fig.~\ref{fig:sec_2}a.

For  non-transparent materials, Magnetic Resonance Imaging (MRI) can obtain contrast-rich images within the bulk~\cite{stannarius:17}. MRI reveals the edges of particles by mapping the distribution of NMR-active nuclei in liquids (or, in exceptional cases also in gases) within a sample. Two requirements have to be met. First, it is necessary to have a liquid  constituent within the sample, since solids typically do not yield useful MRI signals. Second, the liquid needs to have NMR-active nuclei. There are a large number of NMR-active isotopes. While phosphorus, fluorine or  $^{13}$C enriched carbon would work in principle, in practice all commercial MRI scanners work at the proton resonance frequency, i.e. they are tuned to detect the $^1$H nuclei in the sample. For this reason, the most common methods for achieving contrast are to use water- or oil-containing particles (e.g. seeds or synthetic capsules), or to coat/embed solid particles in a hydrogen-containing fluid. In these cases, a conventional tomograph can achieve submillimeter spatial resolution. Samples of several dozen centimeters can be handled in large MRI tomographs, medical scanners, or wide-bore scanners.   

Just as for RIMS, MRI can provide a 3D image or \emph{tomogram} of a sample, from which particle orientations and shapes may be retrieved. In some situations, it suffices to record a single  2D slice of the sample in order to track the dynamics. If so, both  RIMS and MRI can provide single slices at a faster data-collection rate (kHZ). 

Another technique we discuss in this Focus Issue is X-ray tomography\cite{weis:17}. Its advantages are superior spatial resolution and the ability to work with a much more diverse set of materials, including even pasta (see Fig.~\ref{fig:sec_2}b,c.) In contrast to visible light, X-rays will penetrate most granular samples with an intensity that decays exponentially with distance into the material. For a broad range of materials, the X-ray contrast is large enough to provide images. A major drawback, is that the aquisition of even a simple cross-sectional view in general requires the same amount of time as the full 3D image.
 This makes the technique comparatively slow. A compromise is to take 2D transmission images (radiograms) by placing the granular sample between an X-ray source and a camera equipped with a scintillator. In this case, the X-ray signal is integrated across the full sample. 

In computed tomography (CT), a large number (typically $\sqrt{2}$ times the width of the image in pixels) of radiograms is collected while rotating the sample around an axis perpendicular to the beam direction. From this sequence of images, a 3D representation of the sample is reconstructed using an algorithm called the inverse Radon transformation~\cite{buzug:08}. Each voxel represents the  X-ray absorption coefficient within a volume element at the corresponding location in the sample. Since different materials have different absorption coefficients, this information can serve as the contrast to detect the boundaries of particles, and thus allows for identifying the position, shape, and orientation of all granular particles within the tomogram \cite{weis:17}.

The time needed to acquire an X-ray tomogram depends on the photon flux from the source. There are two types of sources: classical X-ray tubes and synchrotrons. X-ray tubes are comparatively low cost, allowing the production of turnkey, tabletop tomography setups with resolutions down to the sub-micron range. Such setups can even be assembled by scientists themselves \cite{athanassiadis:14}. However, X-ray tubes  provide a low photon flux and thus require 10 minutes to several hours to generate a single 3D image.  Synchrotrons are several orders of magnitude brighter than X-ray tubes, thus allowing the acquisition of tomograms at rates up to  several per second. However, because they are large user facilities, they require significant lead time starting with an application for beam time. Moreover, their field of view (FOV) is typically limited to less than a cubic centimeter. 
 
\begin{table*}
     \begin{tabular}{l || c | c | l | l | l  }
         \hline
      Technique & 2D/3D   & spatial     & maximal  & range of &comments  \\ 
                &       &  resolution & framerate& materials&  \\ 
   
         \hline
         \hline
         Disassembly, excavation & 3D  & better than particle size & 1 per hours &practically all & destructive\\
		\hline
         Index-matching (RIMS) & 2D and 3D  & 10 $\mu$m/voxel & 1 per min. & transparent & \\
         \hline
         MRI &2D and 3D & $\approx 100~\mu$m/voxel& 10 per sec.&liquids containing $^1$H & \\
         \hline
         Optical imaging & 2D &  1 $\mu$m/pixel & 10 per millisec. & broad &\\
         \hline
         X-ray tomo  (X-ray tube)& 3D &   1 $\mu$m/voxel & 1 per 10 min. & broad & \\
         \hline
         X-ray tomo  (synchrotron)& 3D &  1 $\mu$m/voxel & 5 per sec. & broad  & small FOV\\
     \end{tabular}
    \caption{Techniques for obtaining particle coordinates, typical values}
    \label{table:positions}
\end{table*}

All of the imaging methods discussed above provide a 2D or 3D image of a granular sample. The algorithmic analysis steps required to quantify particle properties from such images are in principle generic, and can in many cases be applied to images of any imaging method. However, since every imaging method comes with its own specific noise and artifacts, an algorithmic analysis of image data usually begins with a method-dependent step for  denoising and removing artifacts. 

Even with perfect image denoising and artifact removal, there remain significant fundamental challenges in image analysis. A denoised image will be a 2D or 3D set of gray values. The gray value of each pixel/voxel can variously represent the  amount of directly reflected light, the concentration of excited fluorescent dye, a density of spins,  or an X-ray absorption coefficient. This could be measured for either the particle, or for its surrounding medium. Either way, the number of pixels/voxels is limited, and the gray values are drawn from a bounded set of integers. Thus, both the spatial resolution and the image contrast gradients are always finite. This limitation yields a fuzziness on the precise contour of every imaged particle. This is especially detrimental in images of dense particulate media: the fuzziness limits our ability to recognize where particles are exactly located, whether two neighboring objects are touching, or even whether they are in fact separate entities.

A natural first technique is to simply apply a gray value threshold, but this is often insufficient for separation and identification of particles. Instead, an extensive number of algorithms have been developed, often with significant input from the field of computer science, which allow for distinguishing  particles both from the background and from each other~\cite{weis:17}. In addition, there are a number of post-processing methods to detect the location of a particle surface, for which the shape does not have to be known \textit{a priori}~\cite{dijksman:17,vlahinic:14}. 

After the extraction of  detailed shape information about the location of the outer boundary of each particle, further analysis of the experimental data becomes possible. Common techniques include examining nearest neighbor distributions~\cite{weis:17}, particle tracking and velocimetry (see Sec.~\ref{sec:displacements}), contact force measurements (see Sec.~\ref{sec:forces}) and more intricate measures that characterize the packing structure, such as Minkowski tensors~\cite{weis:17}.

\section{Measuring particle displacements \& velocities \label{sec:displacements} }

Measuring velocities starts with first measuring the displacements of objects or patterns between two snapshots separated by a time interval. These measurements can then be accumulated into either a velocity field (for a Eulerian viewpoint) or particle-trajectories (Lagrangian viewpoint), according to the needs of the researcher. 

In making displacement measurements, two  fundamental limitations must be considered. First, the precision with which the position of the objects is known, to be improved according to the techniques described in Section~\ref{sec:positions}. Second, what choice of frame rate can be obtained using the chosen particle-finding method, and whether this timescale is sufficiently well-separated from the dynamics of interest. Several key sources of error inherent in these choices are reviewed in \cite{Xu2004}.

\begin{table*}
     \begin{tabular}{l || c | c | l | l }
         \hline
      Technique & 2D/3D & tracer/   & maximal  & range of \\ 
                &       & all part. & framerate& materials  \\ 
         \hline
         \hline
         DWS & (2 + 1/2)D & all & limited by the photon flux  & multiple scattering \\
         &   & & at high frame rate  & low absorption \\
         \hline
         Index-matching (RIMS)& 2D and 3D & tracer or all & O(10kHz) (tracer); 1/min (all) & transparent \\
     	\hline
         MRI &2D and 3D& tracers or all& $\approx 10$/s& only 
         liquids containing $^1$H \\
         \hline
         PEPT & 3D & tracer& O(kHz) &any\\
         \hline
         PIV & 2D, (2+1)D, 3D & all & Interframe time: O(100 ns)& broad\\
         \hline
         Radar & 3D & tracer & operating frequency, O(GHz) &   contrast of dielectric constant \\
         \hline
         Time-Resolved PIV & 2D & all & O(10kHz)& broad\\
          \hline
         X-ray radiograms & 2D & tracer& 100 Hz (X-ray tube) & density contrast\\
         \end{tabular}
    \caption{Techniques for obtaining particle velocities and displacements}
    \label{table:velocity}
\end{table*}

The frame rate depends on parameters such as the desired resolution (number of pixels/voxels) and the exposure time needed for a sufficient signal-to-noise ratio. Depending on the specific questions of interest, different experimental strategies can be chosen. Innovative methods are often custom-designed for a particular situation. Performance indicators of the different displacement/velocity methods  discussed in this Focus Issue  are summarized in Table~\ref{table:velocity}.

A straightforward method to obtain the Lagrangian trajectories of well-identified particles is to obtain a temporal sequence of their successive positions, from which to  deduce from their full velocities. High-precision particle position measurements play a key role in the success of this method, and identifying single particles is not trivial on its own. One way to track individual particles is to tag a small number of tracers which can be easily distinguished. This can be done with either one or several cameras, to increase spatial or dimensional coverage.  Tracking the centers of the particles is sufficient for studying translational motion.  Rolling or sliding particle motion can be identified only by taking into account further characteristics of the extended particle like form, intensity distribution or tracking marks or patterns on the particles, which will be discussed further in an article of this Focus Issue~\cite{agudo2017detection}.

For particles which are opaque or not index-matched with the surrounding media, velocity measurements are restricted to near-surface flows or to low particle concentrations. Refractive-index matching between fluid and particles allows for the study of deeper layers in granular matter, a technique discussed in one of the articles of the Focus Issue~\cite{dijksman:17}. Other techniques for tracking within the bulk include  Positron Emission Particle Tracking (PEPT)~\cite{parker:17}, microwave radar tracking~\cite{huang:17}, X-ray radiography~\cite{kollmer:17}, or Magnetic Resonance Imaging (MRI)~\cite{stannarius:17}. Tracer particles (single radioactively-labeled in the case of PEPT, high dielectric constant for radar tracking, steel spheres for X-ray radiography, or NMR-active or spin-labeled nuclei for MRI) are then embedded inside the material and the system is imaged while subjected to some excitation or loading. The knowledge of the complete trajectory of one or a few particles in 3D is then possible. In such methods, slow frame rates are not an impediment to the tracking of particles as the particles are well-separated in space. In the case of MRI, magnetization relaxation of labeled tracers may limit the total duration of the dynamics studied. Using PEPT \cite{parker:17} a single particle can be followed almost indefinitely, and in a well-mixed system can provide representative information on the entire phase space of this type of particle. In other systems, the trajectory of a single tracer is not always representative of the whole velocity field, so that full-field methods are preferred.

\begin{figure}[htbp]
\centering
\includegraphics[width=\linewidth]{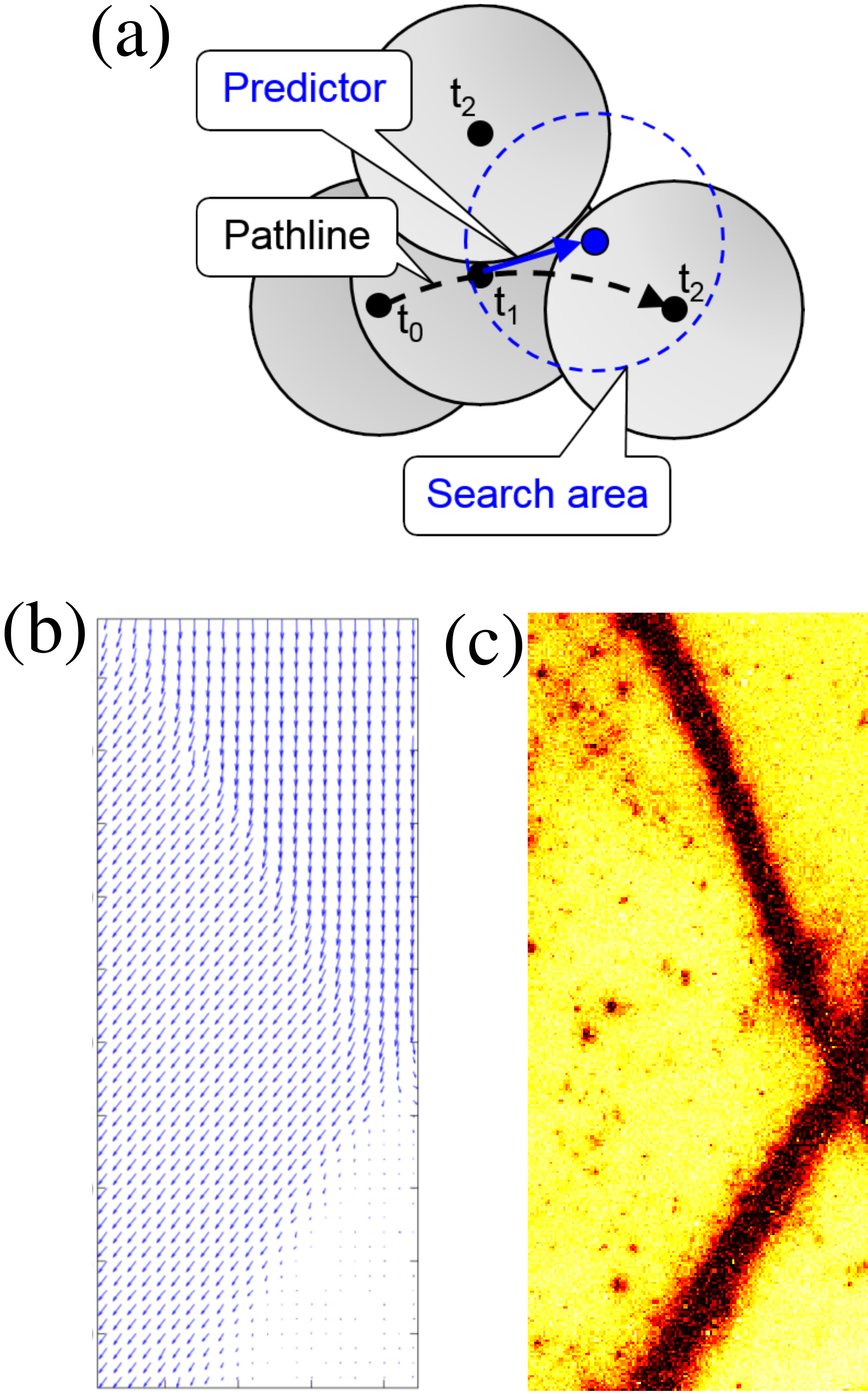}
\caption{(a) Sketch for particle matching in PTV using a predictor from previous displacement. This method allows for more accurate tracking of the particles than a mere nearest-neighbor algorithm. The same displacement field obtained by two different methods (b,c) during a biaxial test after the failure of the material. (b) Displacement field obtained by tracking the reflection of the glass beads at the surface of the sample (PTV). (c) Deformation field obtained by computing correlation between two successive speckle images obtained from the coherent light backscattered by the material (DWS). Yellow: low deformation ($\lesssim 10^{-5}$), black: large deformation ($\gtrsim 10^{-4}$). A description of the experiment and results can be found in~\cite{amon:17}
\label{fig:sec_3}
} 
\end{figure}

In order to follow an assembly of indistinguishable grains, particle tracking velocimetry (PTV) can be achieved by matching particle positions across sequential images~\cite{tuyen2012single, cierpka2013higher, gao2013review}. To discriminate identical particles in subsequent images, the displacement must be smaller than typically half the spacing between particle centers,  which requires that the frame rate is sufficiently large compared to the dynamics of the particles \cite{Xu2004}.
For particle-matching between different frames, various algorithms can be employed. The principle behind tracking algorithms is shown in Figure~\ref{fig:sec_3}(a) and an example of a displacement field obtained using PTV is shown in Figure~\ref{fig:sec_3}(b). An important issue is obtaining sufficiently-accurate positions to identify displacements, while still operating at high enough frame to capture the dynamics. This trade-off is made differently for different systems. In general, 3D scans are used for studying slower dynamics, while fast dynamics require fast cameras aimed at a single slice of the material or a Q2D system. 

Where it is not possible to identify individual particles, it is still possible to measure mean displacements (or velocity/deformation fields) using  cross-correlation techniques like particle image velocimetry (PIV)~\cite{adrian2011particle, raffel2007particle,cierpka2012particle}. These techniques work by subdividing individual frames into small interrogation areas. For these areas, the mean displacement is determined by cross-correlation between subsequent frames. A sub-pixel resolution can be achieved by an appropriate choice of imaging conditions and the size of the interrogation area. Using more than one camera, stereoscopic or volumetric studies can be carried out.

Two entirely different classes of methods allow for the measurement of velocities without the step of identifying particles. 
The first is MRI, for which a useful variant utilizes a time-varying magnetic field as a probe. This allow for the velocities of the MRI-active particles to be deconvolved from the response signal~\cite{Fukushima1999}. 
The second class of methods are all based on the scattering of waves, as described below. 

One variant of the second class is to identify the phase shift of the scattered wave with radio detection and ranging (radar) systems, because it is determined by the relative distance between the scatterer and the observer. Radar methods can be implemented using electromagnetic wavelengths from visible light (LIDAR) through microwave and radio waves ($>10$ m) depending on the applications~\cite{Skolnik2001}. Tracers with a good contrast with the dielectric constant of the surrounding particles are then needed. For a moving target, a radar system compares the phase shift of the microwave being transmitted-to and scattered-from the object to get the position (moving target indicator radar). Alternatively, it uses the Doppler effect to obtain the velocity (Doppler radar)~\cite{Skolnik2008}. The main advantage of radar particle tracking is the good time resolution.  For continuous wave radar systems, the time resolution is only limited by the analog-to-digital converter. Methods for using this technique to obtain the displacement and consequently the instantaneous velocity of a tracer are described in an article of this Focus Issue \cite{huang:17}.

The other variant is to embrace the multiple-scattering limit, which leads to a loss of the information contained in the propagating wave~\cite{born:17}. Terahertz methods are not yet capable of measuring displacements or velocities. In the case of visible light and transparent grains, multiple scattering is unavoidable. Nevertheless, for illumination using coherent light, relative displacements can be obtained from the resulting speckle images arising from the interference of the numerous waves which have performed random walks in the system, a method called Diffusing-Wave Spectroscopy (DWS)~\cite{Weitz1993,Viasnoff2002}. As with any interferometric method, minute relative displacements can be detected and typically deformations of the order of $10^{-5}$ can be obtained. In the case of back-scattering, most of the paths explore only a small volume in the vicinity of the illuminated plane. This means that  spatially-resolved displacement maps, with a resolution of typically a few bead diameters, can be obtained as a representation of the deformation field in a thin layer close to the plane of visualization. An example of a map of deformation obtained by this method is shown in Figure~\ref{fig:sec_3}(c). These techniques are reviewed in an article of this Focus Issue ~\cite{amon:17}.

Finally, as discussed in the introduction, acoustic echo~\cite{Manneville2004} or Doppler measurement~\cite{Takeda1986} can be used to measure velocities in the limit of single scattering of the acoustic waves. The usefulness of such methods is rapidly limited by multiple scattering in granular piles~\cite{Jia1999}. Still, velocity fields have been recently obtained in dense suspensions using high-speed ultrasound imaging~\cite{Han2016}.


\section{Measuring inter-particle forces \label{sec:forces}}


\begin{figure}
\includegraphics[width=0.42\textwidth]{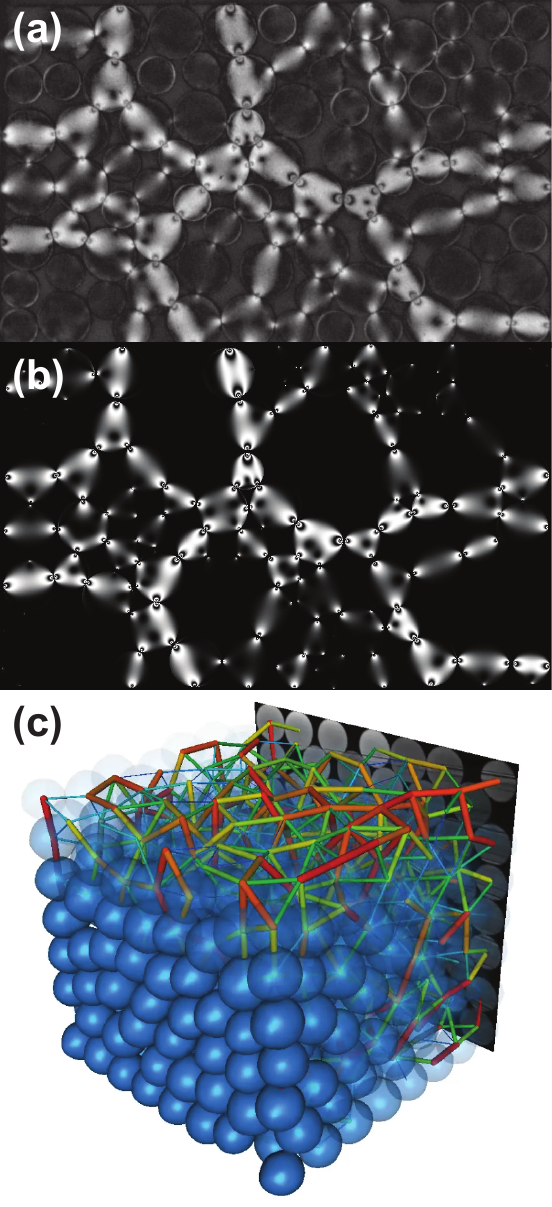}
\caption{(a) Image of force chains obtained using darkfield photoelastic measurements in the setup described in~\cite{ren2013}. The bright areas indicate the regions where contact forces modulate the photoelastic response of the disks. The quantitative extraction of contact forces works by finding the best set of contact forces to ``fit'' the bright pattern. (b) A computer generated fit to the data shown in  (a). (c) Image of 3D interparticle forces obtained on frictionless hydrogel particles, from the experiment described in~\cite{brodu}.}
\label{f:forces}
\end{figure}

While rheological measurements \cite{coussotrheo} can provide the relationship between the bulk stress, strain, and strain-rate dynamics, these  macroscopic tests do not provide insight into how the macroscopic behavior of a particulate material emerges from the interaction laws of the constituent particles. In particular, as can be seen from Fig.~\ref{f:forces}, the patterns of force transmission can be highly heterogeneous. 
To close this loop, it is necessary to track the internal (particle-scale) deformation of a sample, as discussed in Section~\ref{sec:positions} and \ref{sec:displacements}. However, in order to link microscale information to macroscopic stresses, the only route is through a coarse grained formalism such as Irving-Kirkwood \cite{irvingkirkwood}. This formalism requires knowledge of not only particle positions, but also interparticle forces. Therefore measuring \emph{forces between particles} becomes necessary to completely understand particulate materials. 

The requirement to measure forces between particles in such particulate materials presents the experimentalist with great challenges. While Debye theory can predict heat capacities by making simplifying assumptions about the number of modes in a crystal lattice, and a Lennard-Jones interaction potential  suffices to predict the elastic behavior of most solids, granular materials require a more detailed model. First, the interaction laws between particles are generically dissipative, hysteretic, rate dependent, and not pairwise additive \cite{brodu_mc, Hohler}. Second, the disordered nature eliminates the concept of linear response, such that ``small deformation'' analyses are of little use in understanding realistic materials. 

Although it has been possible to obtain stress information at the boundary of a 3D system \cite{eric_02,mueth_98}, true understanding only comes if the spatiotemporal distribution of contact forces is also measured in the bulk of a particulate material, as it is often inhomogeneous and anisotropic. The complex distribution of interparticle forces thus requires the experimentalist to measure not only the already-complex interaction forces between particles, but also their spatial structure in generally opaque or at highly scattering media. The daunting task of performing complex force measurements in the bulk of a particulate material was steadily solved by the work of successive generations of physicists and engineers. 

Photoelasticity, the rotation of polarized light due to the stresses inside a birefringent material,  has a century-old history \cite{dhartog,Dantu1957,Liu1995} of measuring stress and strain distributions within both solids and granular materials. However, until the pioneering work of Behringer et al. \cite{Majmudar2005} in which both normal and tangential interparticle forces were first measured, photoelastic studies were limited to the study of stress fields, and not the interparticle contact forces themselves. The quantitative photoelastic method has evolved tremendously and is now in advanced development across multiple labs around the world. These modern, quantitative methods are described here in a Focus Issue article \cite{daniels:17}, as well as in the original theses \cite{dwhthesis,tsmthesis}. A remaining challenge is that these methods are only suitable for Q2D studies if quantitative information is required. 

Building on the development of conventional confocal microscopy tools, Brujic et al.~\cite{brujic2003} showed that it was possible to measure contact forces in the bulk of a 3D microemulsion (frictionless droplets) using fluorescent dyes. Since then, the emulsion measurement technique has grown to include a range of 3D force measurement techniques, including at larger granular length scales for frictionless particles \cite{2006_science_zhou, brodu, Desmond2013}. In all cases, the central challenge is to measure the size of the contacts, which proceeds using similar techniques to those described in Section~\ref{sec:positions}. As described in a Focus Issue article \cite{dijksman:17}, such measurements are now possible while simultaneously performing mechanical tests under realistic strains.  Due to the experimental and computational complexity of such studies, we present an overview of (partly published) experimental and computational methods used in the study of packings of hydrogel spheres. Similarly, X-ray tomography of non-transparent, frictional particles can provide normal forces for sufficiently soft particles \cite{Saadatfar2012}.

In addition, several other techniques not covered by the Focus Issue bear mentioning since they provide contact force measurements for hard, frictional particles for which deformations are not directly observable.  Using a combination of x-diffraction and X-ray tomography, Hurley et al.\cite{Hurley2016} have successfully measured inter-particle forces for a small sample of non-transparent particles under compression, and at the single-particle level. Chen et al. \cite{Chen2006} have shown that it is possible to use two-photon excitation to make pressure measurements within single ruby particles, based on shifts in the fluorescence spectrum. It is an area of future research to use these methods to understand frictional responses through both the normal and tangential forces.

\section{Acquiring other properties \label{sec:otherprops}}

\begin{figure*}
\centering
\includegraphics[width=0.8\textwidth]{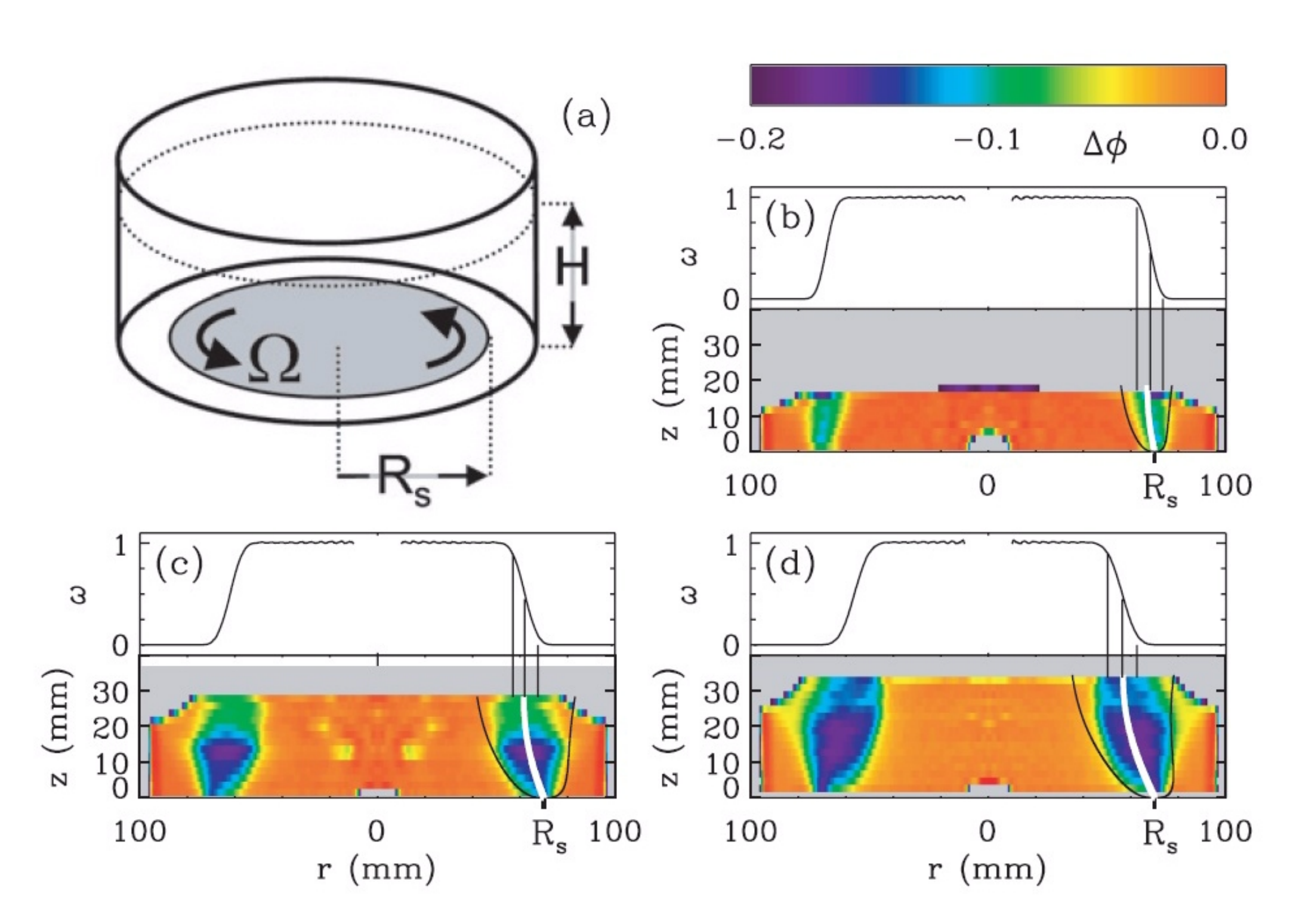}
\caption{Visualization of shear dilatancy by means of MR imaging. In this study, a disk in the bottom of a cylindrical container filled with poppy seeds is continuously rotated until shear dilatancy reaches a stationary value (a). The drop of packing fraction $\Delta\Phi$ was determined from MRI intensity profiles. Subfigures (b-d) show the dilatancy profiles for different filling heights (bottom) together with the profiles of the rotation velocity at the granular surface (top). The shear zone edges are indicated by black lines. Reproduced with permission from EPL 84, 38001 (2008). Copyright 2008 IOP Publishing.
\label{fig:sakaie}
}
\end{figure*}

The imaging of particles within granular media is done with an interest in measuring intensive properties such as the mean particle size, size distribution, packing density, or structural measures. The analysis of large quantities of particles in order to calculate size distributions or packing structures are limited by the generic issues discussed in Sec.~\ref{sec:challenges}. 
When further investigations do not require positions and shapes of the individual particle, investigation methods directly sensitive to such intensive material properties can be an effective alternative. We restrict the discussion to methods which are used in statistical physics analysis of granular media. Methods that were developed for process technologies like sieving or elutriation are beyond the scope of this Focus Issue.

\paragraph{Mean size and size distribution:} The fundamental task is to provide a measurement of the spatial extent of the particles in the system. This can be done by imaging a sufficiently large number of particles,  with sufficient resolution, using any of the methods listed in Sec.~\ref{sec:positions}.  However, it may take an excessively long time to image the ensemble and process the data (separating, identifying, and fitting each one) using one of the imaging methods optimized for opaque 3D materials \cite{weis:17}. Specialized automated setups can image and process tens of thousands of particles in seconds, if disassembly of the sample and ex-situ characterization of the particles are possible \cite{List2011}. Particle sizes between a few micrometers and a few millimeters can be resolved by imaging.

An alternative to imaging is small-angle light scattering. This method relies on the multiple-scattering of light intensity by particles in a well-collimated light beam. The relative intensity changes are sensitive to the particle size, but only when the particles are not much larger than the wavelength of the light used. Consequently, specialized instruments are required to resolve the intensity changes by such scattering \cite{Xu2000}. Recently, terahertz radiation with 87~$\mu$m wavelength has been applied to macroscopic granular particles, which has lowered the instrumental demands to a manageable range \cite{Born2014}. 

In general, scattering methods can be applied to any particle material. Mean particle size and particle size distributions can be determined from single measurements on particle ensembles. Nonetheless, the samples must be disassembled and heavily diluted in order to reach the independent-scattering regime. In practice, this leads to measurement times no faster than the automated imaging setups, and only radii of equivalent spheres are obtained \cite{Xu2000}. Particles sizes between hundred nanometers and roughly a millimeter can be analyzed with scattering.

\paragraph{Packing density and packing structure:} For some imaging methods which are sensitive to an extensive property of the sample, it is possible to use them in a way that reports the average packing density. For example, in X-ray radiography  the intensity decreases exponentially with the number of particles in the beam path. Other examples include index matching and MRI (compare Sec.~\ref{sec:positions}).
For index-matched and fluorescent-dyed samples, the intensity of the signal is proportional to the number of particles containing fluorescence markers (or inversely-proportional, if the liquid contains the fluorescence molecules). For MRI, the signal is proportional to the number of particles containing NMR-active liquids. This means that these methods can switch to a mode which records packing density, even when the resolution would be insufficient to resolve individual particles. 
In each case, the signal is normalized by a sample volume to obtain the packing density. These techniques have allowed X-ray radiograms to be used to study density variations in flowing sand \cite{michalowski:84,baxter:89}, dilatancy during fast shear \cite{Desrues2004,kabla:09,kollmer:17}, or the formation of a jet after a sphere impact into a granular medium \cite{royer:05,homan:15}. MRI has been further used to quantify the evolution of the local particle density {\it e.g.}~in colloid transport \cite{Baumann:05}, during sedimentation  \cite{Ovarlez:12,Morosov:13,Zhu:13,Harich:16}, or in smooth granular shear \cite{Sakaie2008} (see Fig.~\ref{fig:sakaie}).

Conventionally, methods that are directly sensitive to the packing structure rely on measuring scattering patterns. The scattering pattern is sensitive to the packing structure only for samples which exhibit single scattering of a certain wave \cite{Feigin1987}. This is frequently the case for colloidal and atomic systems and light or X-ray scattering, but packing density and scattering efficiency of granular media are too large for this technique to work \cite{born:17}. One novel approach which moves the field towards direct structure sensitivity is presented in this Focus Issue. In \cite{born:17}, we show that the spectroscopic transmission of terahertz radiation through granular materials reveals the position of the structure factor peak (the strongest correlation length in the sample). This approach requires that the particle diameter is within the spectral range of the wavelengths used, and that the particle material does not absorb the radiation too much.

\paragraph{Particle contact stability:}
The local stability and the statistics of particle contacts in consolidated packings can be qualitatively probed with sound. Sound waves propagate within granular media along the force chains formed by the discrete particle contacts \cite{Owens2011} and the velocity of the acoustic wave gives indication on the structure of the granular pile~\cite{Lherminier2014}. Scattering and dissipation of the wave occurs at each contact, leading to diffusive transmission of sound and high attenuation. Simulations have shown that the scattering losses are sensitive to the degree of disorder in the contact network \cite{Mouraille2008}. Intensity losses by dissipation are sensitive to friction at the contacts \cite{Jia2004}. Time-of-flight measurements allow separating the contributions to attenuation by scattering and dissipation \cite{Jia2004}. The momentum and kinetic energy associated with the sound wave are often sufficient to induce rearrangements in granular media with low confining pressures, which can itself be used to probe rigidity and local unjamming in granular media \cite{Wildenberg2015,Lidon2016}. Further investigations of the individual contributions of dissipation, scattering, order and mechanical impact by sound waves might lead to new methods to probe statistics of the contact network in granular media without having to image individual contacts.

\section{Conclusion and Outlook}

This Focus Issue reviews methods for acquiring microscopic particle properties, and for connecting them to the macroscopic physics of granular media. A variety of methods are presented, utilizing  electromagnetic waves ranging from $\gamma$-rays to radio waves. These methods provide information in the form of images, scattering, tomographic reconstruction, and the tracking of phase shifts. Each approach and probe has specific demands on sample material, instrumental investments, and computational efforts, and offers different sensitivity and spatiotemporal resolution. This introductory article aims to  assist the reader in selecting the most appropriate techniques for their particular research. 

To consider the success of these methods, it is only necessary to return to the question raised in the initial paragraph of this paper: how do  individual grains contribute to the intermittent flow of sand while walking on the beach? In spite of the latest advances in methods presented here,  a comprehensive answer to such a question remains beyond reach of even the most advanced contemporary techniques. 

As such, the methods presented here demand future developments to improve both quality and precision. The continued closing of the gap between lab and beach can be deduced from the remaining limitations discussed in this paper. We expect most beneficial improvements will arise in the following area:
\textit{(1)} efficient tracking of both the translation (easier) and rotation (harder) of particles; 
\textit{(2)} techniques for investigating packings of irregular particles;
\textit{(3)} reconstructing 3D force distributions; 
\textit{(4)} improved computational capability and efficiency.
In particular, the simultenous tracking of rotational motion of many particles is still in its infancy. Working with irregular particles may move studies away from the ``spherical cow'' paradigm towards more realistic sand grains. The development of low-cost 3D printers is already fostering rapid interest in  more complex shaped particles \cite{gramabydesign}. Approaches to the reconstruction of forces in 3D packings of hard particles have been demonstrated by fluorescence imaging and  X-ray diffraction. 

Item \textit{(4)} is particularly noteworthy, as it might seem that Moore's Law will take care of this problem without interference from granular scientists. However, each of the other three items on the list significantly increase the amount of data to be handled. Tracking rotation and translation at the same time doubles the amount of data per particle, working with irregular particles requires better shape detection algorithms and keeping track of more descriptive variables, and the reconstruction of 3D force distributions requires new detection and fitting algorithms as well as an increase in the number of parameters (contacts and force vectors) stored in the final dataset. As the spatiotemporal resolution of tomographic methods improve, the amount of data grows as a cubic function. As such, data science initiatives and algorithmic advances will be key to the success of the field.

\bibliography{rsi_in_one_plus_arxiv} 

\end{document}